\documentstyle[prd,aps,floats]{revtex}
\begin{document}

\input epsf

\draft
%
%
\twocolumn[\hsize\textwidth\columnwidth\hsize\csname
@twocolumnfalse\endcsname
\preprint{SUSSEX-AST 97/1-3, SUSX-TH-97-002, CU-TP-809, CAL-625, 
astro-ph/9702006}
\title{On the instability of the one-texture universe}
\author{Xuelei Chen\cite{xcaddress}}
\address{Department of Physics, Columbia University, 538 West
120th Street, New York, New York 10027}
\author{Mark Hindmarsh\cite{mhaddress}}
\address{Centre for Theoretical Physics, University of Sussex, Falmer, 
Brighton BN1 9QH, United Kingdom}
\author{Marc Kamionkowski\cite{mkaddress}}
\address{Department of Physics, Columbia University, 538 West
120th Street, New York, New York 10027}
\author{Andrew R. Liddle\cite{aladdress}}
\address{Astronomy Centre, University of Sussex, Falmer, 
Brighton BN1 9QH, United Kingdom}
\maketitle
\begin{abstract}
The one-texture universe, introduced by Davis in 1987, is a
homogeneous mapping of a scalar field with an $S^3$ vacuum into a
closed universe. It has long been known to mathematicians that such
solutions, although static, are unstable. We show by explicit construction
that there are four degenerate lowest modes which are unstable, 
corresponding to
collapse of the texture towards a single point, in the case where
gravitational backreaction is neglected. We discuss the instability
timescale in both static and expanding space-times; in the latter case
it is of order of the present age of the universe, suggesting that,
though unstable, the one-texture universe could survive to the
present.  The cosmic microwave background constrains the initial
magnitude of this unstable perturbation to be less than ${\cal
O}(10^{-3})$.
\end{abstract}
\pacs{PACS numbers: 98.80.Hw\\[3pt]
 Preprint: SUSSEX-AST 
96/1-3, SUSX-TH-97-002, CU-TP-809, CAL-625, astro-ph/9702006}

\vskip2pc]

\section{Introduction}

A texture is a topological defect which arises in scalar field
theories with a spontaneously broken global symmetry, when the vacuum
manifold $M$ has a non-trivial third homotopy group $\pi_3$. The
simplest such manifold is the three-sphere $S^3$. In a spatially flat
universe, the texture is known to be unstable to collapse, and indeed
this property is crucial for the texture model of structure formation
in the universe \cite{Turok}. The one-texture universe (which predates
and has nothing to do with the texture scenario for structure
formation), introduced by Davis \cite{Davis}, considers instead a
closed universe, itself with $S^3$ topology, and arises when one maps
the vacuum manifold directly onto the configuration space. This gives
a non-trivial homogeneous solution to the equations of motion, the
scalar field possessing an energy density associated with its spatial
gradients.  This scalar-field configuration has an equation of state
$p=-\rho/3$ and gives rise to a term in the Friedmann equation which
scales as $a^{-2}$ with the scale factor $a$ of the universe.
Recently, Kamionkowski and Toumbas \cite{KT} showed that a universe
with a matter density less than unity can be closed (and consistent
with current observations) with the inclusion of matter with
$p=-\rho/3$.

Aware that textures collapse in flat space through shrinking, Davis
considered a particular shrinking ansatz for the one-texture universe,
and showed that it was stable to this particular kind of perturbation
\cite{Davis}. He concluded from this that the one-texture universe was
stable. However, this does not correspond to a complete stability
analysis. Shortly after we began looking at this problem, it was
pointed out to us by Durrer that in fact the answer was already in the
mathematics literature \cite{math}, predating even the original Davis
paper. The one-texture universe is {\em unstable}.  Indeed, there is
no non-constant stable map from any compact manifold into $S^3$
\cite{math}.

However, the results in the mathematics literature apply, strictly
speaking, only to a static universe.  It is well known that the
expansion of the universe slows the growth of density perturbations
and similarly that the expansion of a deflagration bubble slows the
growth of hydrodynamic instabilities in the bubble wall \cite{kf}.  It
is therefore of interest to see whether the growth of instabilities in
the one-texture universe are slowed by the expansion of the universe,
and to evaluate the timescale for instability.

In this paper, we consider the instability by explicit
construction. We show that there are four degenerate lowest modes which are
unstable, and analyze their growth rate in both static and 
expanding space-times.

In the next Section, we present some preliminaries regarding the
closed universe and the scalar field theory.  In Section III, we first
review the texture scalar field configuration.  We then find an
unstable mode and analyze its growth rate in a static universe, in a
radiation-, matter-, texture-, and cosmological-constant dominated
universe, and in a universe with comparable matter and texture energy
densities.  The growth rate in an expanding universe is slowed
compared with the exponential rate in a static universe, and during
inflation, the instability is frozen.  The instability timescale is
comparable to the age of the universe.  In Section IV, we perform a
general perturbation analysis and find that there are four degenerate
modes including the one found in Section III.  There are also six 
zero modes corresponding to rotations and translations.  All other higher
modes are stable and decay in an expanding universe.
We conclude in Section V and argue
that isotropy of the cosmic microwave background only constrains the
magnitude of the initial unstable mode to be less than ${\cal
O}(10^{-3})$.

\section{Preliminaries}

The space-time metric for a closed universe can be written as
\begin{eqnarray}
ds^2 & = & a^2(\tau) \left[ -d\tau^2 + d\xi^2 + \sin^2\xi \, \left( 
	d\theta^2 + \sin^2 \theta \, d\phi^2 \right) \right] \,, 
	\nonumber \\
& & \hspace*{0.2cm} \mbox{with~~} 0 \leq \xi < \pi, 0 \leq \theta < \pi,
	0 \leq \phi < 2\pi \,,
\end{eqnarray}
where $\tau$ is the conformal time and $a(\tau)$ the scale factor. 

We consider a theory of four real scalar fields with an O(4) global
symmetry, spontaneously broken to O(3) by a suitable potential.  The
vacuum manifold for the scalar field is therefore $S^3$. If we
consider the long-wavelength modes, the massive degree of freedom
(i.e.~the $\sigma$-meson in a linear $\sigma$-model) would not be
excited, and the dynamics are approximately those of a non-linear
sigma model with this target space.  The action is therefore
\begin{equation}
S = \frac{v^2}{2} \int d^4x \sqrt{|g|} \, G_{AB}(X) \partial_\mu
	X^A(x)\partial_\nu X^B(x) g^{\mu\nu}(x) \,,
\label{e:action}
\end{equation}
where $X^A$ ($A=1,2,3$) are coordinates on the target space $M$, which 
has metric $G_{AB}$, and $v$ is the vacuum expectation of the scalar field.  
We will also use polar coordinates for $M$, namely
$\Xi$, $\Theta$, $\Phi$, so that $G_{AB} = {\rm 
diag}(1,\sin^2\Xi,\sin^2\Xi\sin^2\Theta)$. Their ranges are as with $\xi$, 
$\theta$ and $\phi$ respectively.

The equations of motion for $X^A$ are 
\begin{eqnarray}
\frac{1}{\sqrt{|g|}}\, 
{\partial_\mu} \left(\sqrt{|g|} \, g^{\mu\nu}
	\partial_\nu X^A\right) &+& \nonumber\\
& & \hspace*{-3cm} \Gamma^A_{BC}(X)\partial_\mu X^B\partial_\nu 
	X^C g^{\mu\nu}(x) = 0 \,,
\label{e:EOM}
\end{eqnarray}
where $\Gamma^A_{BC}$ are the Christoffel symbols of the metric $G_{AB}$ 
on $M$. Solutions to these equations were called 
{\em harmonic maps} by Misner \cite{Mis78}. 

In our stability analysis we will need to consider the eigenmodes of
small perturbations around static solutions $X^A=f^A$ to these
equations, for which we will write $X^A=f^A + \epsilon n^A$. It is
possible to define a covariant derivative for $n^A$, by
\begin{equation}
{n^A}_{;\mu} = \partial_\mu n^A + (\Gamma^A_{BC}\partial_\mu f^C)n^B \,.
\end{equation}
The equation for linear perturbations around $f^A$ may then be written
\begin{equation}
\label{e:pert}
{{n^A}_{;\mu}}^{;\mu} + (R^A_{CBD}\partial_\mu f^C\partial^\mu f^D)n^B 
	= 0 \,,
\end{equation}
where $R^A_{CBD}$ is the Riemann curvature of the sigma model target 
space $M$.

Before solving the perturbation equation in its generality, we
describe the one-texture universe and examine a simple perturbation
around it. The lowest eigenvalue solution of this perturbation equation
will turn out to be one of a set of four degenerate unstable modes.

\section{The One-Texture Universe and a Perturbation}

\label{sec3}

The trivial solution has the symmetry breaking pointing in the same
direction (say $\Xi = \Theta = \Phi = 0$) at all points of space. By
contrast, the one-texture universe corresponds to the solution
\cite{Davis}
\begin{equation}
\Xi = \xi, \quad \Theta = \theta, \quad \Phi = \phi \,.
\end{equation}
That this is a solution is clear once we realize that $e_i^A \equiv
\partial_i f^A =\delta_i^A$, where $i$ and $j$ are spatial indices.
Then it follows that $G^{AB} = \delta_i^A\delta_j^B g^{ij}$, and
Eq.~(\ref{e:EOM}) becomes the well-known geometric identity
\begin{equation}
\frac{1}{\sqrt{|G|}}\, {\partial_B} \left(\sqrt{|G|}\,G^{BA}\right) 
	+ \Gamma^A_{BC}G^{BC} = 0 \,.
\end{equation}

We now consider a linear perturbation to the radial component which is
a function of $\xi$ only. This is the simplest guess for an unstable
mode, and as we shall see turns out to be correct.  We therefore let
$\Xi = \xi + \delta(\xi,\tau)$.  An equation of motion for
$\delta(\xi,\tau)$ can be found by plugging directly into
Eq.~(\ref{e:pert}).  It is easier to vary the action, which for a
perturbation of this form is
\begin{eqnarray}
     S \propto \int\,&& d\tau\, d\xi\, a^2\, \sin^2\xi\,  \Bigl[
     -(\partial \Xi/\partial \tau)^2+ (\partial \Xi/\partial
     \xi)^2 \\
     && + 2 \sin^2\Xi\, \csc^2 \xi \Bigr].
\label{e:specialaction}
\end{eqnarray}
Doing so, we find 
\begin{eqnarray}
\left[ \frac{d^2}{d\xi^2}  + 2 \frac{\cos \xi}{\sin \xi} \frac{d}{d\xi} +
	4 - \frac{2}{\sin^2 \xi} \right] \delta(\xi,\tau) & = & 
	\nonumber \\
 & & \hspace*{-3cm} \left[ \frac{d^2}{d\tau^2} + 2 \frac{\dot{a}}{a}
	\frac{d}{d\tau} \right] \delta(\xi,\tau) \,,
\end{eqnarray}
where overdot is a derivative with respect to the conformal time
$\tau$. As usual, we look for separable solutions $\delta(\xi,\tau) =
\delta(\xi) f(\tau)$. Let us concentrate first on the spatial
eigenmodes, which satisfy
\begin{equation}
\left[ \frac{d^2}{d\xi^2} + 2 \frac{\cos \xi}{\sin \xi} \frac{d}{d\xi} +
	4 - \frac{2}{\sin^2 \xi} \right] \delta(\xi) = -\omega^2 
	\delta(\xi) \,.
\label{e:spert}
\end{equation}
This can in fact be brought into Schr\"{o}dinger equation form by a
change of independent variable to $u(\xi) = \delta(\xi) \sin \xi$,
yielding
\begin{equation}
-\frac{d^2u}{d\xi^2} +\frac{2}{\sin^2 \xi} \, u = \left( \omega^2 + 5
	\right) \, u \,.
\end{equation}
The first two unnormalized eigenmodes are $u_1(\xi) = \sin^2 \xi$ and
$u_2(\xi) = \sin^2 \xi \cos \xi$, with eigenvalues $-1$ and 4
respectively.  These form the first two elements of a series
\begin{equation}
u_n(\xi) = \sin^2 \xi \, C^{(2)}_{n-1}(\cos\xi) \quad (n\geq1) \,,
\end{equation}
where $C^{(\lambda)}_m(t)$ are Gegenbauer polynomials of degree $m$
\cite{GraRyz80}. The eigenvalue corresponding to the $n$th
eigenfunction is $\omega^2 = n(n+2) - 4$.

The stability (or otherwise) now follows from the time eigenfunctions,
which depend on the behavior of the scale factor. They are solutions
to the equation
\begin{equation}
\label{time}
\left[ \frac{d^2}{d\tau^2} + 2 \frac{\dot{a}}{a} \frac{d}{d\tau} \right]
	f(\tau) = -\omega^2 f(\tau) \,,
\end{equation}
which we mention in passing is precisely the equation for the
amplitude of gravitational-wave modes \cite{Grish}, though there the
range of permitted eigenvalues is different.

\subsection{The static universe}

In this case the time eigenfunctions are simply 
\begin{equation}
f(\tau) \propto \exp(\pm i\omega \tau) \,.
\end{equation}
The higher spatial eigenmodes, with positive $\omega^2$, are
oscillatory.  However, the lowest eigenmode has negative $\omega^2$,
and hence corresponds to an exponentially growing instability. This
provides an explicit confirmation of the mathematical result of
Ref.~\cite{math}.

The instability corresponds to the spatial gradients concentrating
towards one of the poles (which one depending on whether the sign of
the perturbation is positive or negative). Presumably by analogy to
the spatially flat case, once the winding is sufficiently concentrated
the texture will be pulled away from its vacuum manifold and the
topological charge disappears.

\subsection{Radiation domination}

A radiation dominated universe has $a(\tau) \propto \tau$. The time
eigenmodes are then
\begin{equation}
f(\tau) \propto \frac{\exp \left( \pm i \omega \tau \right)}{\tau} \,.
\end{equation}
We see that the condition for instability remains $\omega^2 < 0$.
\subsection{Matter domination}

Here $a(\tau) \propto \tau^2$. The time eigenmodes for positive
$\omega^2$ are now given in terms of spherical Bessel functions
\cite{AbSt}
\begin{equation}
f(\tau) =  \alpha_1 \, \frac{j_1(\omega \tau)}{\tau} +
	\alpha_2 \, \frac{y_1(\omega \tau)}{\tau} \,,
\end{equation}
where $\alpha_1$ and $\alpha_2$ are constants. For negative $\omega^2$
these become modified spherical Bessel functions; the linearly
independent mode of the first kind diverges at late times and gives
the instability.

\subsection{Texture-dominated universe without backreaction}

In a texture-dominated closed universe without other matter, $a(\tau) 
\propto \exp (\tau/\sqrt{\Omega_{{\rm T}} -1})$, where the texture energy 
density as a fraction of the critical density, $\Omega_{{\rm T}}$, is a 
constant. The solution to Eq.~(\ref{time}) is of the form
\begin{equation} 
f(\tau) \propto \exp (\beta \tau) \quad ; \quad 
	\beta = -\frac{1\pm\sqrt{1-\omega^{2}(\Omega_{{\rm T}}-1)}} 
	{\sqrt{\Omega_{{\rm T}}-1}} \,.
\end{equation}
For the $\omega^2= -1$ mode, the solutions are just a growing and a
decaying exponential. For positive $\omega^2$, the constant $\beta$ is
complex, with sine and cosine oscillations superimposed on an
exponential decay.

The more interesting solution is one where there is still an
appreciable (non-relativistic) matter density, such as in
Ref.~\cite{KT}. Then, denoting present values with the subscript `0',
the expansion rate is
\begin{equation}
\frac{a(\tau)}{a_0} = \frac{\Omega_{{\rm M},0}}{2(1-\Omega_{{\rm M},0})} 
	\; \left[ \cosh (\alpha_0^{1/2} \tau) -1 \right] \,,
\end{equation}
where the constant $\alpha_0$ is given by
\begin{equation}
\label{alpha0}
\alpha_0 = \frac{1-\Omega_{{\rm M},0}}{\Omega_{{\rm M},0}
	+\Omega_{{\rm T},0}-1}\,,
\end{equation}
where $\Omega_{{\rm M}}$ and $\Omega_{{\rm T}}$ are the energy
densities in matter and texture respectively, in units of the critical
density. For a given choice of $\Omega_{{\rm M},0}$ and $\Omega_{{\rm
T},0}$, the present time $\tau_0$ is found from the requirement
$a(\tau_0) \equiv a_0$, giving
\begin{equation}
\label{tau0}
\tau_0 = \alpha_0^{-1/2} \, \cosh^{-1} \left[ 
	\frac{1-\Omega_{{\rm M},0}/2}{\Omega_{{\rm M},0}/2} \right] \,.
\end{equation}

\begin{figure}
\centering 
\leavevmode\epsfysize=6.5cm \epsfbox{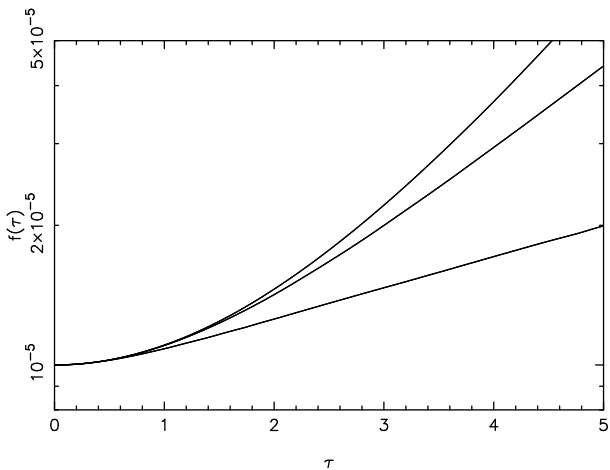}\\ 
\caption[text_time]{\label{text_time} The growth of the unstable 
perturbation in a one-texture universe with matter, shown, from top to 
bottom, 
for $\alpha_0 = 0.1$, $1$, and $10$.} 
\end{figure}

The corresponding equation for the time evolution of the eigenmodes
does not look promising for analytical solution, but is easily solved
numerically.  The solution for the unstable mode is shown in
Fig.~\ref{text_time}, for three values of $\alpha_0$ and for a
perturbation with zero initial velocity and arbitrarily chosen initial
size. The late-time exponential behavior is as predicted from the case
of total texture-domination ($\beta = \sqrt{1+\alpha_0} \, -
\sqrt{\alpha_0}$); the matter-dominated solution applies at early time
but matter domination doesn't last long enough for much growth during
that phase.

Given $\Omega_{{\rm M},0}$ and $\Omega_{{\rm T},0}$, the value of
$f(\tau_0)$ can be found from Fig.~\ref{text_time}.  For example, if
$\Omega_{{\rm M},0}\simeq0.3$ (as some observations indicate), then
$\Omega_{{\rm T},0}\simeq1.8$ if we impose the (perhaps arbitrary)
condition of Ref. \cite{KT} that the cosmic microwave background (CMB)
surface of last scatter be at the antipode.  In this case,
$\alpha_0\simeq0.64$, $\tau_0\simeq3$, and $f(\tau_0)/f(0)\simeq2$. In
general, the time-scale of the unwinding of texture is the same as the
age of the universe, which is not surprising since it is also the
horizon scale.  Therefore, in this low-density closed universe, any
initial perturbations to the texture would have doubled by today. Note
that the radiation-dominated phase preceding the matter dominance
has little effect since the matter--radiation transition occurred at
redshift $z \sim 10^4 $, where $a \sim 10^{-6} a_{0}$. At that epoch
the long wavelength modes had not yet entered the horizon, and so the
growth of perturbations in the texture is negligible.

\subsection{An inflationary universe}

In all the above cases, the instability associated with the $\omega^2
= -1$ mode diverges at late times. In this final case we explore a
slightly different outcome. This occurs in the case of perturbations
in an inflationary universe, which is dominated by vacuum energy or
equivalently by a cosmological constant.  We consider power-law
inflation which has $a \propto t^p$, where $t$ is cosmic time and
$p>1$ is a constant. In conformal time we have
\begin{equation}
a(\tau) \propto (-\tau)^{-p/(p-1)} \quad , \quad -\infty < \tau < 0 \,,
\end{equation}
where late times correspond to $\tau \to 0$ and where exponential inflation 
is recovered in the limit $p \to \infty$. The solution for positive 
$\omega^2$, in terms of fractional-order Bessel functions, is
\begin{equation}
f(\tau) = \alpha_1 \, (-\tau)^{\mu} J_{\mu}(-\omega \tau) +
	\alpha_2 \, (-\tau)^{\mu} Y_{\mu}(-\omega \tau) \,,
\end{equation}
where $\mu = p/(p-1)$ and again $\alpha_1$ and $\alpha_2$ are
constants.  For negative $\omega^2$, the Bessel functions become
modified Bessel functions. In both cases, the solution of the first
kind vanishes at late times, whereas the second kind `freezes out' at
a constant value.

Here the distinction between stability and instability is much
murkier, because of this characteristic type of behavior in
inflationary universes.  The positive $\omega^2$ modes, which would
ordinarily be oscillatory about the stable solution, become frozen
with a displacement away from the stable solution, while the negative
$\omega^2$ mode, that would normally grow to divergence, also becomes
frozen at late times. This interesting behavior has already been
investigated for the case of gravitational-wave perturbations, which
also freeze to a constant value, by Allen \cite{Allen}. He describes
this behavior as `global instability/local stability'; although
globally the de Sitter space becomes more and more distorted as modes
of higher and higher $\omega$ freeze out, the region accessible to any
observer shrinks rapidly during inflation and seems to become more and
more smooth. The same behavior is evident here.

Nevertheless, effectively the nature of the stability is just the same
as before. Note in particular that the $\omega^2 = -1$ mode does not
die away during the late stages of inflation but instead stays fixed;
once radiation or matter domination restarts its growth will begin
again, though on a long timescale since the period of inflation has
stretched it to such a large physical size. Only once the Hubble
length has grown again to be comparable to the size of the closed
universe will the instability set in in earnest.

\section{General Perturbation Analysis}

The general solutions to Eq.~(\ref{e:pert}) are in fact quite
straightforward to find, as they turn out to be vector harmonics on
$S^3$ which allows us to plunder the mathematics literature
\cite{San78}.  Once again, we can use the constancy of $\partial_i
f^A$ to rewrite the perturbation equations as an eigenvalue problem
for the vector $n^A$, as
\begin{equation}
{{n^A}_{;\mu}}^{;\mu} = - a^{-2} \partial_\tau(a^2 \partial_\tau n^A)
	+ {{n^A}_{;C;D}}G^{CD} \,.
\end{equation}
where we use the fact that in the one-texture universe background
$g_{ij} = \delta_i^A\delta_j^B G_{AB}$. Further simplification arises
as the background is a three-sphere, for which
\begin{equation}
R^A_{CBD}G^{CD} = 2 \delta^A_B \,.
\end{equation}
Hence, when we separate the solution, the equation for the spatial
eigenmodes becomes
\begin{equation}
{{n^A}_{;C;D}}G^{CD} + 2 n^A = -\omega^2 n^A \,.
\label{e:eigprob}
\end{equation}
This is an eigenvalue equation for vector harmonics on $S^3$, whose
solutions are known \cite{San78}. They fall into three classes, $A_A$,
$B_A$, and $C_A$, which are odd and even parity divergence-free, and
curl-free respectively. The eigenvalues of the Laplacian for the
curl-free harmonics are $[2 - n(n+2)]$, with $n>0$, while for the
divergence-free harmonics they are $[1 - n(n+2)]$.  Hence our
eigenvalue problem also has three classes of solution classified by
their symmetry properties under $S^3$ coordinate transformations, with
eigenvalues
\begin{eqnarray}
\omega_0^2 &=& n(n+2) -4, \quad n \geq 1 \,, \nonumber\\
\omega_{\pm 1}^2 &=& n(n+2) -3, \quad n \geq 1 \,,
\end{eqnarray}
where the subscript $0$ denotes the curl-free class, and $\pm 1$ the
even and odd parity divergence-free class. The curl-free modes are
those we found in Section \ref{sec3}.

We can now see why the Gegenbauer polynomials made their appearance in
Eq.~(\ref{e:spert}). The curl-free vector harmonics are just the
divergence of the scalar harmonics on $S^3$:
\begin{equation}
C_A(X) = \partial_A Y^{(nlm)}(\xi,\theta,\phi) \,,
\end{equation}
where
\begin{eqnarray}
Y^{(nlm)}(\xi,\theta,\phi) = N_{nl} \sin^l\xi \, 
C^{(l+1)}_{n-l}(\cos\xi)
	Y^{(lm)}(\theta,\phi) \, , \\
-l \le m \le l \le n\,. \nonumber
\end{eqnarray}
(Here, $N_{nl}$ is a normalization factor and $Y^{(lm)}(\theta,\phi)$
are the usual $S^2$ spherical harmonics).  The relation $\partial_t
C^{(\lambda)}_n(t) = 2\lambda C^{(\lambda+1)}_{n-1}(t)$ then allows us
to find all the $l=0$ harmonics in the direction $\partial_\xi X^A$,
which amounts to solving for $\delta(\xi)$ in Eq.~(\ref{e:spert}).

Thus we see that the only unstable modes are those with $n=1$ 
curl-free vector harmonics.  There are four of them, and the spherical
one we analyzed in \S III corresponds $l=m=0$; there are also
three anisotropic modes, which are $l=1, m=0, \pm 1$. As they have the 
same eigenvalue, the time eigenfunctions found in Section \ref{sec3} 
apply to each of them. There are also
six zero modes in the divergence-free class, which correspond to rotations 
and translations of the texture solution. The
rest of the modes are stable, and correspond to propagating Goldstone
bosons.

\section{Conclusions}

We have investigated in detail the nature of the instability in the
one-texture universe.  We have found that there are four degenerate
unstable modes,  and evaluated the growth rate in a radiation-, matter-,
and texture-dominated universe, an inflationary universe, and a
universe with similar matter and texture densities.  In a static
universe, the growth of the instability is exponential.  As one may
have expected, the growth of the instability is slowed during matter,
radiation, or texture domination, and it is frozen during an
inflationary epoch.  Since the eigenvalue of the unstable modes is
$-1$, the timescale for the instability is comparable to the age of
the Universe.  For a given texture and matter density, the growth
factor can be obtained explicitly from the curves in
Fig.~\ref{text_time}.  For the case of a closed Universe with
$\Omega_{{\rm M},0}=0.3$ and a texture density which puts the CMB
surface of last scatter at the antipode, any initial unstable
perturbation would have increased by a factor of two.

Causality restricts the unstable modes from growing until the
wavelength---in this case, the curvature radius---comes within
the horizon.  If the texture density is chosen so that the CMB
is at the antipode, then the curvature radius today coincides
with the horizon.  If so, then the instability timescale is
comparable to the age of the Universe.  If, however, the texture
density is much smaller so that the curvature radius greatly
exceeds the horizon today, then the instability timescale will
be much longer than the age of the Universe today.
Strictly speaking, one should also consider the effect of
gravitational backreaction. However, for most of the cases we
consider here, the texture makes a negligible contribution to
the total energy density, so inclusion of backreaction should
not alter our results qualitatively.  For the texture-dominated
case, backreaction may significantly affect the evolution.
However, causality still restricts modes from growing until
their wavelengths come within the horizon.  Therefore, we do not
expect backreaction to alter our conclusions, although it may
change the instability timescale by factors of order unity.
Similarly, we neglect the
effect of fluctuations in the matter, since they are coupled to the
texture only via gravitation. If these matter fluctuations are
correlated or anti-correlated to the fluctuations in the texture field
initially, as is often assumed in the texture scenario of structure
formation \cite{PST}, then the amplitude of these fluctuations should
be comparable to those of the texture field, and thus their effect
would be second order. If these fluctuations are uncorrelated to the
texture fluctuation, it might happen that their amplitude is much
greater than that of the texture, but in this case it is merely a
random fluctuation on the space-time background, and we would expect
that they would not affect the evolution of long wavelength modes of
texture.

So, what does this analysis tell us about the requirements for the
homogeneity of the initial texture configuration?  First of all, we
note that all modes are stable, except for the long-wavelength $n=1$ 
curl-free modes and for the divergence-free zero modes which correspond
to rotations and translations. 
{}From the results in
Section III, the modes with $\omega^2 > 0$ will decay with the
expansion of the Universe, so one does not require stringent
constraints on the general homogeneity of the initial texture. Since
the instability timescale for the unstable modes is comparable to
the age of the Universe, the initial inhomogeneity of the texture
should be comparable to or less than the inhomogeneity of the Universe
today. These inhomogeneities can induce perturbations in the metric
and further affect the matter distribution, both of which can produce
anisotropy in CMB. The isotropy of the CMB should therefore place an
upper limit to the acceptable magnitude of this perturbation in the
initial scalar-field configuration. Although we have not done a
complete analysis of CMB anisotropies induced by this instability, it
should give rise to a dipole anisotropy on the sky. 
The $n=1$ modes have either $l=1$ or $l=0$.  The $l=1$ modes are obviously 
dipole.  The $l=0$ mode is isotropic to an observer at the origin, but
not to other observers. Indeed, if we make a translation of the origin, the
$S^3$ spherical harmonics with same $n$ but different $l$ will mix.
Thus for a general observer, it appears
anisotropic with a dipole pattern. Given that the CMB dipole is ${\cal
O}(10^{-3})$, this constrains the magnitude of the initial $\omega^2
=-1$ perturbation to be less than roughly this value.  If the CMB
dipole can confidently be aligned with the gradient of the local
density field to, say 10\%, then the constraint to the initial
magnitude of the $\omega^2 =-1$ perturbation should be an order of
magnitude smaller.

We therefore conclude that except for the $\omega^2 =-1$ modes,
inhomogeneities in the scalar-field configuration will decay, so in
some sense, the one-texture universe does not require extraordinarily
peculiar initial conditions.  One only requires that the magnitude of
the lowest-eigenmode perturbation be less than ${\cal O}(10^{-3})$.
Of course, more precise conclusions regarding the implications for the
one-texture universe will have to await a more complete theory of its
origin.

\acknowledgments

We thank John Barrow, Ruth Durrer and Erick Weinberg for discussions. 
M.H.~was supported by PPARC and A.R.L.~by the Royal Society.
This work was supported at Columbia by D.O.E. contract
DEFG02-92-ER 40699, NASA NAG5-3091, and the Alfred P. Sloan
Foundation.  A.R.L.~thanks Columbia University for its
hospitality during a visit, and M.H., M.K.~and A.R.L.~thank the
University of Uppsala/NORDITA for support during a visit to
Sweden.



\begin{references}
\bibitem[\clubsuit]{mhaddress} Electronic address: 
m.b.hindmarsh@sussex.ac.uk
\bibitem[\heartsuit]{aladdress} Electronic address: a.liddle@sussex.ac.uk
\bibitem[\diamondsuit]{xcaddress} Electronic address: 
xuelei@phys.columbia.edu
\bibitem[\spadesuit]{mkaddress} Electronic address: kamion@phys.columbia.edu
\bibitem{Turok} N. Turok, Phys. Rev. Lett. {\bf 63}, 2625 (1989).
\bibitem{Davis} R. L. Davis, Phys. Rev. D {\bf 35}, 3705 (1987).
\bibitem{KT} M. Kamionkowski and N. Toumbas, Phys. Rev. Lett. {\bf 77},
	597 (1996).
\bibitem{math} Y. L. Xin, Duke Math. J. {\bf 47}, 609 (1980); P. F.
	Leung, Lecture Notes in Mathematics {\bf 949}, 122
	(Springer-Verlag, Berlin, 1982); Y. Ohnita, T\^ohoku Math.
	Journ. {\bf 38}, 259 (1986).
\bibitem{kf} M. Kamionkowski and K. Freese,
        Phys. Rev. Lett. {\bf 69}, 2743 (1992).   
\bibitem{Mis78} C. W. Misner, Phys. Rev. D {\bf 18}, 4510 (1978).
\bibitem{GraRyz80} I. S. Gradshteyn and I. M. Ryzhik, 
	{\em Table of Integrals, Series, and Products}  
	(Academic Press, New York, 1994).
\bibitem{Grish} L. P. Grishchuk, Zh. Eksp. Teor. Fiz. {\bf 67}, 823 
	(1974) [Sov. Phys. JETP {\bf 40}, 409 (1974)].
\bibitem{AbSt} M. Abramowitz and I. A. Stegun, {\em Handbook of 
	Mathematical Functions} (Dover Publications, NY, 1972).
\bibitem{Allen} B. Allen, Phys. Rev. D {\bf 37}, 2078 (1988).
\bibitem{San78} V. D. Sandberg, J. Math. Phys. {\bf 19}, 2441 (1978).
\bibitem{PST} See for example U. Pen, D. N. Spergel and N. Turok,
	Phys. Rev. D. {\bf 49}, 692 (1994).
\end{references}
\end{document}